\documentclass[12pt]{article}

\usepackage{scicite}

\usepackage{times}

\newenvironment{sciabstract}{%
\begin{quote} \bf}
{\end{quote}}

\newcounter{lastnote}
\newenvironment{scilastnote}{%
\setcounter{lastnote}{\value{enumiv}}%
\addtocounter{lastnote}{+1}%
\begin{list}%
{\arabic{lastnote}.}
{\setlength{\leftmargin}{.22in}}
{\setlength{\labelsep}{.5em}}}
{\end{list}}

\newcommand{\beq}{\begin{equation}}
\newcommand{\eeq}{\end{equation}}
\newcommand{\beqa}{\begin{eqnarray}}
\newcommand{\eeqa}{\end{eqnarray}}
\newcommand{\beqar}{\begin{eqnarray*}}
\newcommand{\eeqar}{\end{eqnarray*}}

\def\bi{\begin{itemize}}
\def\ei{\end{itemize}}
\def\be{\begin{equation}}
\def\ee{\end{equation}}
\def\bea{\begin{eqnarray}}
\def\eea{\end{eqnarray}}
\def\ben{\begin{eqnarray*}}
\def\een{\end{eqnarray*}}

\def\>{\rangle}
\def\<{\langle}

\newcommand{\1} I 

\newcommand{\ket}[1]{| #1 \rangle}

\def\*{\star}

\def\tilde{\widetilde}
\def\bar{\overline}

        \def\cB{{\cal B}}
\def\cC{{\cal C}}
        \def\cE{{\cal E}}

\def\cG{{\cal G}}

\def\cS{{\cal S}}        

\def\cV{{\cal V}}        

        \def\cZ{{\cal Z}}

\title{Correcting Quantum Errors with Entanglement}

\author{Todd Brun$^{1\ast}$, Igor Devetak$^1$, Min-Hsiu Hsieh$^1$ \\
\normalsize{$^{1}$Electrical Engineering Department, University of Southern California,} \\
\normalsize{Los Angeles, CA 90089, USA}\\
\\
\normalsize{$^\ast$To whom correspondence should be addressed; E-mail:  tbrun@usc.edu.}}

\date{ }

\begin{document}

\maketitle


\begin{sciabstract}
We show how entanglement shared between encoder and decoder can
simplify the theory of quantum error correction.  The
entanglement-assisted quantum codes we describe do not require the
dual-containing constraint necessary for standard quantum error
correcting codes, thus allowing us to ``quantize'' all
of classical linear coding theory. In particular, efficient modern
classical codes that attain the Shannon capacity can be made into
entanglement-assisted quantum codes attaining the hashing bound
(closely related to the quantum capacity). For systems without large
amounts of shared entanglement, these codes can also be used as
catalytic codes, in which a small amount of initial entanglement
enables quantum communication.
\end{sciabstract}

Entanglement plays a central role in quantum information processing.
It enables the teleportation of quantum states without physically
sending quantum systems\cite{BBCJPW93}; it doubles the capacity of
quantum channels for sending classical information\cite{BW92}; it is
known to be necessary for the power of quantum
computation\cite{BCD02,JL03}. We show how shared entanglement
provides a simpler and more fundamental theory of quantum error
correction.

The theory of quantum error correcting codes was established a decade ago as the primary tool for fighting decoherence in quantum computers and quantum communication systems.  The first nine-qubit single error-correcting code was a quantum analog of the classical repetition code, which stores information redundantly by duplicating each bit several times \cite{PS95}.  Probably the most striking development in quantum error correction theory is the use of the stabilizer formalism\cite{CRSS97,DG97thesis,DG98,NC00}, whereby quantum codes are subspaces (``code spaces'') in Hilbert space, and are specified by giving the generators of an abelian subgroup of the Pauli group, called the stabilizer of the code space. Essentially, all QECCs developed to date are stabilizer codes.  The problem of finding QECCs was reduced to that of constructing classical dual-containing quaternary codes\cite{CRSS97}.  When binary codes are viewed as quaternary, this amounts to the well known Calderbank-Shor-Steane construction\cite{Ste96,CS96}.  The requirement that a code contain its dual is a consequence of the need for a commuting stabilizer group.  The virtue of this approach is that we can directly construct quantum codes from classical codes with a certain property, rather than having to develop a completely new theory of quantum error correction from scratch.  Unfortunately, the need for a self-orthogonal parity check matrix presents a substantial obstacle to importing the classical theory in its entirety, especially in the context of modern codes such as low-density parity check (LDPC) codes \cite{MMM04}.

Assume that the encoder Alice and decoder Bob have access to shared
entanglement. We will argue that in this setting every quaternary
(or binary) classical linear code, not just dual-containing codes, can be
transformed into a QECC, and illustrate this with a particular
example.  If the classical codes are not dual-containing, they
correspond to a set of stabilizer generators that do not commute;
however, if shared entanglement is an available resource, these
generators may be embedded into larger, commuting generators, giving
a well-defined code space. We call this the entanglement-assisted
stabilizer formalism, and the codes constructed from it are
entanglement-assisted QECCs (EAQECCs).

\paragraph*{Standard stabilizer formalism.}
The power of the stabilizer formalism comes from the clever use of
group theory. Let $\Pi$ denote the set of Pauli operators
$\{I,X,Y,Z\}$, and let $\Pi^n=\{I,X,Y,Z\}^{\otimes n}$ denote the
set of $n$-fold tensor products of single-qubit Pauli operators.
Then $\Pi^n$ together with the possible overall factors $\pm 1, \pm
i$ forms a group $\cG_n$ under multiplication, the $n$-fold Pauli
group.  Here are a few useful properties of the $n$-fold Pauli
group: (a)  every element of $\cG_n$ squares to $\pm I_n$ (plus or
minus the identity); (b)  any two elements of $\cG_n$ either commute
or anti-commute; (c)  every element of $\cG_n$ is unitary; and (d)
elements of $\cG_n$ are either Hermitian or anti-Hermitian. The
connection of $\cG_n$ to error correction is straightforward:  the
elements of $\cG_n$ can be identified as possible sets of errors
that might affect a quantum register of $n$ qubits.

Suppose $\cS$ is an abelian subgroup of $\cG_n$.
We define the stabilizer code $\cC(\cS)$ associated with $\cS$ to be
\[
\cC(\cS)=\{\ket{\psi}:M\ket{\psi}=\ket{\psi}, \forall M\in \cS\}.
\]
The code $\cC(\cS)$ is the subspace fixed by $\cS$, so $\cS$ is
called the stabilizer of the code. In other words, the code space is
the simultaneous $+1$ eigenspace of all elements of $\cS$. For an
$[[n,k]]$ stabilizer code, which encodes $k$ logical qubits into $n$
physical qubits, $\cC(\cS)$ has dimension $2^k$ and $\cS$ has
$2^{n-k}$ elements\cite{NC00}.  We should notice that for group $S$
to be the stabilizer of a nontrivial subspace, it must satisfy two
conditions:  the elements of $\cS$ commute, and $-I_n$ is not in
$\cS$.  (This second condition implies that all elements of $\cS$
are Hermitian, and hence have eigenvalues $\pm1$.)

A group $\cS$ can be specified by a set of independent generators,
$\{M_i\}$. These are elements in $\cS$ that cannot be expressed as
products of each other, and such that each element of $\cS$ can be
written as a product of elements from the set.  If an abelian
subgroup $\cS$ of $\cG_n$ has $2^{n-k}$ distinct elements up to an overall phase,
then there are $n-k$ independent generators. The benefit of using
generators is that it provides a compact representation of the
group; and to see whether a particular vector $\ket\psi$ is stabilized by
a group $\cS$, we need only check whether $\ket\psi$ is stabilized by
these generators of $\cS$.

Suppose $\cC(\cS)$ is a stabilizer code, and the quantum register is
subject to errors from an error set $\cE = \{E_a\} \subset \cG_n$.
How are the error-correcting properties of $\cC(\cS)$ related to the
generators of $\cS$?  First, suppose that $E_a$ anti-commutes with a
particular stabilizer generator $M_i$ of $\cS$.  Then
\[
M_i E_a\ket{\psi}=-E_a M_i\ket{\psi}=-E_a\ket{\psi}.
\]
$E_a\ket\psi$ is an eigenvector of $M_i$ with eigenvalue $-1$, and
hence must be orthogonal to the code space (all of whose vectors
have eigenvalue $+1$).  As the error operator $E_a$ takes the
code space of $\cC(\cS)$ to an orthogonal subspace, an occurrence of
$E_a$ can be detected by measuring $M_i$.  For each generator $M_i$
and error operator $E_a$, we can define a coefficient $s_{i,a}
\in\{0, 1\}$ depending on whether $M_i$ and $E_a$ commute or
anti-commute:
\[
M_i E_a=(-1)^{s_{i,a}}E_a M_i.
\]
The vector $\bar{s}_a=(s_{1,a},s_{2,a},\cdots,s_{n-k,a})$ represents
the syndrome of the error $E_a$.  In the case of a nondegenerate
code, the error syndrome is distinct for all $E_a \in\cE$, so that
measuring the $n-k$ stabilizer generators will diagnose the error
completely.  However, a uniquely identifiable error syndrome is not
always required for an error to be correctable.

What if $E_a$ commutes with the generators of $\cS$?  If $E_a \in
\cS$, we do not need to worry, because the error does not corrupt the
space at all.  The real danger comes when $E_a$ commutes with all
the elements of $\cS$ but is not itself in $\cS$.  The set of
elements in $\cG_n$ that commute with all of $\cS$ is the
centralizer $\cZ(\cS)$ of $\cS$.  If $E \in \cZ(\cS)-\cS$, then $E$
changes elements of $\cC(\cS)$ but does not take them out of
$\cC(\cS)$. Thus, if $M \in \cS$ and $\ket{\psi}\in \cC(\cS)$, then
\[
ME\ket{\psi}=EM\ket{\psi}=E\ket{\psi}.
\]
Because $E \not\in \cS$, there is some state of $\cC(\cS)$ that is not
fixed by $E$.  $E$ will be an undetectable error for this code.
Putting these cases together, a stabilizer code $\cC(\cS)$ can
correct a set of errors $\cE$ if and only if $E_a^\dagger E_b\in \cS
\cup (\cG_n-\cZ(\cS))$ for all $E_a, E_b \in \cE$.

\paragraph*{Entanglement-assisted stabilizer codes.}
We will now illustrate the idea of the entanglement-assisted stabilizer
formalism by an example.  We know from the previous paragraph that a
stabilizer code can be constructed from a commuting set of operators
in $\cG_n$.  What if we are given a non-commuting
set of operators? Can we still construct a QECC? Let $\cS$ be the
group generated by the following non-commuting set of operators: \be
\begin{array}{ccccc}
\label{setM}
M_1= & Z & X & Z & I   \\
M_2= & Z & Z & I & Z   \\
M_3= & X & Y & X & I   \\
M_4= & X & X & I & X
\end{array}
\ee It is easy to check the commutation relations of this set of
generators: $M_1$ anti-commutes with the other three generators,
$M_2$ commutes with $M_3$ and anti-commutes with $M_4$, and $M_3$
and $M_4$ anti-commute.  We will begin by finding a different set of
generators for $\cS$ with a particular class of commutation
relations.  We then relate $\cS$ to a group $\cB$ with a
particularly simple form, and discuss the error-correcting
conditions using $\cB$.  Finally, we relate these results back to
the group $\cS$.

To see how this works, we need two lemmas.  (See the supporting online
material for proofs.)  The first lemma shows
that there exists a new set of generators for $\cS$ such that $\cS$
can be decomposed into an ``isotropic'' subgroup $\cS_I$ generated by a set of
commuting generators, and a ``symplectic'' subgroup $\cS_S$ generated by a set of
anti-commuting generator pairs \cite{FCY04}.

\noindent {\it Lemma 1.} Given any arbitrary subgroup $\cV$ in
$\cG_n$ that has $2^{m}$ distinct elements up to overall phase,
there exists a set of $m$ independent generators for $\cV$ of the form
$\{\bar{Z}_1,\bar{Z}_2,\cdots,\bar{Z}_\ell,\bar{X}_1,\cdots,\bar{X}_{m-l}\}$
where $m/2 \leq \ell \leq m$, such that $[ \bar{Z}_i , \bar{Z}_j] =
[ \bar{X}_i , \bar{X}_j] = 0$, for all $i,j$; $[ \bar{Z}_i ,
\bar{X}_j] = 0$, for all $i \neq j$; and $\{ \bar{Z}_i , \bar{X}_i
\}= 0$,  for all $i$. Here $[A,B]$ is the commutator and $\{A,B\}$
the anti-commutator of $A$ with $B$.  Let
$\cV_I = \langle\bar{Z}_{m-\ell+1},\cdots,\bar{Z}_\ell\rangle$ denote the
isotropic subgroup generated by the set of commuting generators,
and let
$\cV_S = \langle\bar{Z}_1,\cdots,\bar{Z}_{m-\ell},\bar{X}_1,\cdots,\bar{X}_{m-\ell}\rangle$
denote the symplectic subgroup generated by the set of
anti-commuting generator pairs. Then, with slight abuse of the
notation, $\cV = \langle\cV_I , \cV_S\rangle$ indicates that $\cV$ is generated
by subgroups $\cV_I$ and $\cV_S$.

For the group $\cS$ that we are considering, one such set of
independent generators is \be
\begin{array}{ccccc}
\bar{Z}_1= & Z & X & Z & I   \\
\bar{X}_1= & Z & Z & I & Z   \\
\bar{Z}_2= & Y & X & X & Z   \\
\bar{Z}_3= & Z & Y & Y & X
\end{array}
\ee so that $\cS_S = \langle\bar{Z}_1,\bar{X}_1\rangle$,
$\cS_I = \langle\bar{Z}_2,\bar{Z}_3\rangle$, and $\cS=\langle\cS_I,\cS_S\rangle$.

The choice of the notation $\bar{Z}_i$ and $\bar{X}_i$ is not
accidental: these generators have exactly the same commutation
relations as a set of Pauli operators $Z_i$ and $X_i$ on a set of
qubits labeled by $i$.  Let $\cB$ be the group generated by the
following set:
\be
\begin{array}{ccccc}
Z_1= & Z & I & I & I  \\
X_1= & X & I & I & I  \\
Z_2= & I & Z & I & I  \\
Z_3= & I & I & Z & I  \\
\end{array}.
\ee
From the previous lemma, $\cB = \langle\cB_I,\cB_S\rangle$, where
$\cB_S = \langle Z_1,X_1\rangle$ and $\cB_I = \langle Z_2,Z_3\rangle$. Therefore, groups $\cB$ and
$\cS$ are isomorphic, which is denoted as $\cB\cong\cS$.  We can
then relate $\cS$ to the simpler group $\cB$ by the following lemma
\cite{BFG05}:

\noindent {\it Lemma 2.} \,\, If $\cB\cong\cS$, then there exists a
unitary $U$ such that for all  $B\in\cB$  there exists an
$S\in\cS$ such that $B = U S U^{-1}$ up to an overall phase.

As a consequence of this lemma, the error-correcting power of
$\cC(\cB)$ and $\cC(\cS)$ are also related by a unitary
transformation.  In what follows, we will use $\cB$ to discuss the
error-correcting conditions, and then translate the results back to
$\cS$.

What is the code space $\cC(\cB)$ described by $\cB$? Because $\cB$
is not a commuting group, the usual definition of $\cC(\cB)$ does
not apply, as the generators do not have a common eigenspace.
However, by extending the generators, we can find a new group that
is commuting, and for which the usual definition of code space can
apply; the qubits of the codewords will be embedded in a larger
space.  Notice that we can append a $Z$ operator at the end of
$Z_1$, an $X$ operator at the end of $X_1$, and an identity at the
end of $Z_2$ and $Z_3$ to make $\cB$ abelian: \be
\begin{array}{ccccc|c}
Z_1'= & Z & I & I & I & Z \\
X_1'= & X & I & I & I & X \\
Z_2'= & I & Z & I & I & I \\
Z_3'= & I & I & Z & I & I
\end{array}
\ee We assume that the four original qubits are possessed by Alice
(the sender), and the additional qubit is possessed by Bob (the
receiver) and is not subject to errors.  Let $\cB_e$ be the extended
group generated by $\{Z_1',X_1',Z_2',Z_3'\}$.  We define the code
space $\cC(\cB)$ to be the simultaneous $+1$ eigenspace of all
elements of $\cB_e$, and we can write it down explicitly in this
case:
\be
\cC(\cB)=\{\ket{\Phi}^{AB}\ket{0}\ket{0}\ket{\psi}\},
\ee
where $\ket{\Phi}^{AB}$ is a maximally entangled state shared
between Alice and Bob, and $\ket{\psi}$ is an arbitrary single-qubit
pure state.  Because entanglement is used, this is an EAQECC.
We use the notation $[[n,k;c]]$ to denote an EAQECC that
encodes $k$ qubits into $n$ qubits with the help of $c$ ebits.
(Sometimes we will write $[[n,k,d;c]]$ to indicate that the ``distance'' of the
code is $d$, meaning it can correct at least $\lfloor\frac{d-1}{2}\rfloor$ errors.)  The
number of ebits $c$ needed for the encoding is equal to the number
of anti-commuting pairs of generators in $\cB_S$. The number of
ancilla bits $s$ is equal to the number of independent generators in
$\cB_I$. The number of encoded qubits $k$ is equal to $n-c-s$, and
we define the rate of the EAQECC to be $(k-c)/n$. Therefore,
$\cC(\cB)$ is a $[[4,1;1]]$ EAQECC with zero rate:  $n=4$, $c=1$,
$s=2$ and $k=1$. Note that the zero rate does not mean that no
qubits are transmitted by this code!  Rather, it implies that a
number of bits of entanglement is needed that is equal to the
number of bits transmitted.  In general, $k-c$ can be positive,
negative, or zero.

Now we see how the error-correcting conditions are related to the
generators of $\cB$. We saw that if an error $E_a \otimes I^B$
anti-commutes with one or more of the operators in
$\{Z_1',X_1',Z_2',Z_3'\}$, it can be detected by measuring these
operators.  This will only happen if the error $E_a$ anti-commutes
with one of the operators in the original set of generators
$\{Z_1,X_1,Z_2,Z_3\}$, as the entangled bit held by Bob is assumed
to be error-free. Alternatively, if  $E_a\otimes I^B \in \cB_e$, or
equivalently if $E_a \in \cB_I$, then $E_a$ does not corrupt the
encoded state. In this case we call the code degenerate. Altogether,
$\cC(\cB)$ can correct a set of errors $\cE_0$ if and only if
$E_a^\dagger E_b \in \cB_I \cup (\cG_4-Z(\cB))$ for all
$E_a,E_b\in\cE_0$.

With this analysis of $\cB$, we can go back to determine the
error-correcting properties of our original stabilizer $\cS$.  We
can construct a QECC from a nonabelian group $\cS$ if entanglement
is available, just as we did for the group $\cB$.  We add extra
operators $Z$ and $X$ to make $S$ abelian as follows: \be
\begin{array}{ccccc|c}
\label{setS}
\bar{Z}_1'= & Z & X & Z & I & Z  \\
\bar{X}_1'= & Z & Z & I & Z & X  \\
\bar{Z}_2'= & Y & X & X & Z & I  \\
\bar{Z}_3'= & Z & Y & Y & X & I
\end{array}
\ee where the extra qubit is once again assumed to be possessed by
Bob and to be error-free. Let $\cS_e$ be the group generated by the
above operators.  Since $\cB \cong \cS$, let $U$ be the unitary from
Lemma 2.  Define the code space $\cC(\cS)$ by
$\cC(\cS)=U^{-1}(\cC(\cB))$, where the unitary $U$ is applied only
on Alice's side.  This unitary $U$ can be interpreted as the
encoding operation of the EAQECC defined by $\cS$.  Observe that the
code space $\cC(\cS)$ is a simultaneous eigenspace of all elements
of $\cS_e$.  As in the analysis for $\cC(\cB)$, the code $\cC(\cS)$
can correct a set of errors $\cE$ iff
$E_a^\dagger E_b \in \cS_I \cup (\cG_4-Z(\cS))$ for all $E_a,E_b\in\cE$.

The algebraic description is somewhat abstract, so let us translate
this into a physical picture.  Alice wishes to encode a single
($k=1$) qubit state $\ket\psi$ into four ($n=4$) qubits, and
transmit them through a noisy channel to Bob.  Initially, Alice and
Bob share a single ($c=1$) maximally entangled pair of qubits---one
ebit.  Alice performs the encoding operation $U$ on her bit
$\ket\psi$, her half of the entangled pair, and two ($s=2$) ancilla
bits.  She then sends the four qubits through the channel to Bob.
Bob measures the extended generators $\bar{Z}_1', \bar{X}_1',
\bar{Z}_2'$, and $\bar{Z}_3'$ on the four received qubits plus his
half of the entangled pair.  The outcome of these four measurements
gives the error syndrome; as long as the error set satisfies the
above requirement, Bob can correct the error and decode the
transmitted qubit $\ket\psi$.

We have worked out the procedure for a particular example, but any
EAQECC will function in
the same way.  The particular parameters $n,k,c,s$ will vary
depending on the code.  It should be noted that the first example of
entanglement-assisted error correction produced a $[[3,1,3;2]]$
EAQECC based on the $[[5,1,3]]$ standard QECC
\cite{Bow02}. Our construction differs in that it is completely
general and, more important, eschews the need for commuting
stabilizers.

\paragraph*{Construction of EAQECCs from classical quaternary codes.}
We will now examine the $[[4,1;1]]$ EAQECC given above, and show
that it can be derived from a classical non-dual-containing
quaternary $[4,2]$ code. This is a generalization of the well-known
construction for standard QECCs\cite{CRSS98}.

First, note that this $[[4,1;1]]$ code is non-degenerate, and can
correct an arbitrary one-qubit error. (Therefore the distance $d$ of
the code $\cC(\cS)$ is 3.) This is because the $12$ errors $X_i,
Y_i$ and $Z_i$, $i = 1, \dots, 4$, have distinct non-zero error
syndromes. $X_i$ denotes the bit flip error on $i$-th qubit, $Z_i$
denotes the phase error on the $i$-th qubit, and $Y_i$ means that both a
bit flip and phase flip error occur on the $i$-th qubit.  It
suffices to consider only these three standard one-qubit errors,
because any other one-qubit error can be written as a linear
combination of these three errors and the identity.

Next, we define the following map between the Pauli operators and
elements of GF(4), the field with four elements:
\[
\begin{array}{|c|c|c|c|c|} \hline
\Pi &  I & X & Y & Z \\  \hline
GF(4) &  0 & \overline{\omega} & 1 & \omega \\ \hline
\end{array}
\]
Note that under this map, addition in GF(4) corresponds to multiplication of the Pauli operators, up to an overall phase.  So multiplication of two elements of $\cG_n$ corresponds to addition of two $n$-vectors over GF(4), up to an overall phase.

The set of generators $\{M_i\}$ given in Eq.~(\ref{setM}) is mapped to the matrix $\tilde{H}_4$:
\be
\tilde{H}_4 =
\left(\begin{array}{cccc}
\omega & \bar{\omega}  & \omega & 0  \\
\omega & \omega  & 0 & \omega  \\
\bar{\omega} & 1 & \bar{w} & 0  \\
\bar{\omega} & \bar{\omega} & 0 & \bar{\omega}
\end{array} \right).
\ee Examining the matrix $\tilde{H}_4$, we see that it can be
written \be \label{CtoQ} \tilde{H}_4=\left(\begin{array}{c} \omega H_4 \\
\bar{\omega}H_4 \end{array}\right) ,
\ee
where $H_4$ is the parity-check matrix of a classical $[4,2,3]$ quaternary code
whose rows are not orthogonal, and 3 is the minimum distance between codewords:
\be H_4 = \left(\begin{array}{cccc}
1 & \omega  & 1 & 0  \\
1 & 1  & 0 & 1
\end{array} \right).
\ee

We get a $[[4,1,3;1]]$ EAQECC from a classical $[4,2,3]$ quaternary
code. This outperforms the best 4-bit self-dual QECC currently
known, which is $[[4,0,2]]$\cite{CRSS98}.  This connection between
EAQECCs and quaternary classical codes is quite general\cite{SOM}.
Given an arbitrary classical $[n,k,d]$ quaternary code, we can use
Eq.~(\ref{CtoQ}) to construct a non-degenerate $[[n,2k-n+c,d;c]]$ EAQECC.
The rate becomes $(2k-n)/n$ because the $n-k$ classical parity checks
give rise to $2(n-k)$ quantum stabilizer generators.
(The complete details of this construction, along with rigorous
proofs of its performance can be found in the supporting online
materials.)

\paragraph*{Discussion.}
Our entanglement-assisted stabilizer formalism enables us to
construct QECCs from arbitrary classical
quaternary codes without the dual-containing constraint. The
simplification and unification that occurs when entanglement
assistance is allowed is an effect well known in the context of
quantum Shannon theory\cite{BSST01,DHW03}.

The better the classical quaternary code is, the better the
corresponding EAQECC will be. Searching for good quantum codes now
becomes the problem of searching for good classical codes, which has
been extensively studied and is well understood.  Efficient modern
codes, such as Turbo codes \cite{BGT93} or LDPC codes \cite{RG63thesis} whose
performance approaches the classical Shannon limit, can now be used
to construct corresponding quantum codes.

There are two interesting properties of EAQECCs constructed from
Eq.~(\ref{CtoQ}). A classical quaternary code that saturates the
Singleton bound will give rise to a quantum code saturating the
quantum Singleton bound.  To see this, assume that the $[n,k,d]$ classical
quaternary code saturates the classical Singleton bound;
that is, $n-k\geq d-1$. The corresponding $[[n,2k-n+c,d;c]]$ quantum
code then saturates
\[
n-(2k-n)=2(n-k)\geq 2(d-1),
\]
which is the quantum Singleton bound\cite{KL97}.

Another feature of EAQECC is that a classical quaternary code that
achieves the Shannon bound will give rise to a quantum code that
achieves the ``hashing'' limit on a depolarizing
channel\cite{BDSW96}.  Let the rate (in base 4) $R_C$ of a $[n,k,d]$ quaternary
code meeting the Shannon bound of the quaternary symmetric channel be
\[
R_C = C_4(f) = 1 - ( H_4(f) + f \log_4 3 ),
\]
where $f$ is the error probability and $H_b(f) = - f \log_b f - (1-f) \log_b (1-f)$
is the entropy in base $b$. Then the rate (in base 2) $R_Q$
of the corresponding $[[n,2k-n+c,d;c]]$ EAQECC is
\[
R_Q = 2R_C-1 = 1 - ( H_2(f) + f \log_2 3 ) ,
\]
which is exactly the hashing bound on a depolarizing channel.  The
hashing bound is a lower bound on the closely related quantum channel
capacity.  It was previously achieved only by inefficient random coding
techniques \cite{BDSW96}.

The use of an EAQECC requires an adequate supply
of entanglement.  However, these codes can be useful even if there
is not a large amount of pre-existing entanglement, by turning an
EAQECC into a catalytic QECC (CQECC).  The
idea here is simple.  Suppose the EAQECC has parameters $n,k,c$.
Using $c$ bits of pre-existing entanglement, Alice encodes some of
the qubits she wishes to transmit, plus one bit each from $c$
maximally entangled pairs that she prepares locally.  After her $n$
bits have been transmitted to Bob, corrected and decoded, Bob has
received $k-c$ qubits, plus $c$ new bits of entanglement have been
created.  These can then be used to send another $k-c$ bits, and so
on. The idea is that the perfect qubit channel that is simulated by
the code is a stronger resource than pre-existing
entanglement\cite{DHW03}.  It is this catalytic mode of performance
that makes the rate $(k-c)/n$ a reasonable figure of merit for an EAQECC as
described above.  Clearly the $[[4,1,3; 1]]$ code described in this
paper is useless as a catalytic code, though it is perfectly useful
for an entanglement-assisted channel.  To be a useful catalytic
code, an EAQECC must have a positive value of $k-c$.

We have presented EAQECCs in a communication context up to now,
but catalytic codes open the possibility of application to error correction in
quantum computing, where we can think of decoherence as a channel
into the future.  In this case, the ``seed'' resource is not
pre-existing entanglement, but rather a small number of qubits that
are error-free, either because they are physically isolated, or
because they are protected by a decoherence-free subspace or standard QECC.

CQECCs provide great flexibility in designing quantum communication
schemes.  For example, in periods of low usage we can use an EAQECC
in the catalytic mode to build up shared entanglement between Alice
and Bob.  Then in periods of peak demand, we can draw on that
entanglement to increase the capacity.  Quantum networks of the
future can use schemes like this to optimize performance.  In any
case, the existence of practical EAQECCs will greatly enhance the power of quantum
communications, as well as providing a beautiful connection to the
theory of classical error correction codes.

\bibliography{ref4}

\begin{thebibliography}{10}
\bibitem{SOM}
Proofs and the detailed construction are available as supporting material on
  Science Online.
\end{thebibliography}

\bibliographystyle{Science}

\begin{scilastnote}
\item We would like to acknowledge helpful feedback from David Poulin, Graeme Smith and Jon Yard.  TAB acknowledges financial support from NSF Grant No.~CCF-0448658, and TAB and MHH both received support from NSF Grant No.~ECS-0507270.  ID and MHH acknowledge financial support from NSF Grant No.~CCF-0524811 and NSF Grant No.~CCF-0545845.
\end{scilastnote}

\end{document}